\documentclass[aps,pra,twocolumn,superscriptaddress]{revtex4}

\usepackage{amssymb}
\usepackage{graphicx}
\usepackage{amsmath}
\usepackage{xcolor}

\newcommand{\be}{\begin{equation}}
\newcommand{\ee}{\end{equation}}
\newcommand{\ba}{\begin{eqnarray}}
\newcommand{\ea}{\end{eqnarray}}

\begin{document}
\title{Anderson Localization for Very Strong Speckle Disorder}
\author{Michael Hilke}
\email{hilke@physics.mcgill.ca}
\affiliation{Department of Physics, McGill University, Montr\'eal, QC, Canada, H3A 2T8}
\author{Hichem Eleuch}
\email{heleuch@fulbrightmail.org}
\affiliation{Institute for Quantum Science and Engineering,
Texas A$\&$M University, College Station, Texas 77843, USA}
\affiliation{Department of Physics, McGill University, Montr\'eal, QC, Canada, H3A 2T8}

\begin{abstract}
We evaluate the localization length of the wave (or Schr\"odinger) equation in the presence of a disordered speckle potential. This is relevant for experiments on cold atoms in optical speckle potentials. We focus on the limit of large disorder, where the Born approximation breaks down and derive an expression valid in the "quasi-metallic" phase at large disorder. This phase becomes strongly localized and the effective mobility edge disappears.

\end{abstract}

\maketitle

\section{Introduction}
There has been a lot of recent excitement with the advent of new experiments and theories on localization, many years after the seminal work of Anderson in 1958 \cite{bib1,bib1a,bib1b,bib1c,bib1d}. Interesting experiments on impurities of crystals in solids \cite{a6}, porous substrate in superfluid Helium \cite{a7} and roughness of the refractive index in photonic crystals \cite{a8} are other examples of disorder in physical systems. New theoretical concepts have also emerged, such as many-body localization with interesting concepts related to thermalization and entanglement entropy. Recent experiments in cold atoms have found evidence of many-body localization \cite{schreiber,choi,kondov}.

Cold atoms are an important example of quantum particles in a potential \cite{a9,a9b,a9c,a9d}. Understanding the interplay between disorder and interaction is well adapted to cold atoms. in these systems, disorder is often created using speckle patterns \cite{a10}.
Optical Speckle patterns can be generated by a transmission of a laser light through a material surface with a roughness on the scale of an optical wavelength \cite{a11}. This effect is recognizable by granular patterns of intensity and stems from the interference of dephased coherent waves. The speckle phenomenon is not limited to optical fields but is also used in several other applications such as in radar\cite{a12}, ultrasound medical imagery \cite{a13} and spectral analysis of random processes \cite{a14}.

The aim of this work is to study wave localization in strong and very strong speckle disorder, where the Born approximation breaks down. We define weak disorder when $\sqrt{\langle V^2 \rangle}\ll E$, where $V$ is the disorder potential of zero average, $\langle \cdot \rangle$ represents the disorder average and $E$ the energy of the particle. Strong disorder is defined when $\sqrt{\langle V^2 \rangle}\simeq E$ but $V<E$, while very strong disorder is when $\sqrt{\langle V^2 \rangle}\gg E$ and $V(x_{sub})>E$, i.e., there are many sub-regions ($x_{sub}$), where the potential exceeds the energy and there exist semi-bound states and tunneling takes place. Speckle potentials are good examples of a class of potentials, where these distinctions are important. Indeed, for red detuned potentials, $\max(V)$ is bound, but not for blue detuned potentials because of the exponential potential distribution \cite{delande}.

Speckle disorder is characterized by a characteristic correlation length, $\xi$, which depends on the speckle generation. This correlation length can be easily tuned, which makes speckle disorder an ideal system to study the effect of $\xi$ on localization properties. Indeed, experiments showing Anderson localization in cold atom systems have used speckle disorder and found an effective mobility edge in 1D because of the high spatial frequency cutoff of the disorder \cite{a17, billy}. Hence, when the de Broglie wavelength becomes much shorter than $\xi$, the particles become delocalized in this quasi-metallic phase.

Here we show that for large enough speckle disorder the wave remains strongly localized even in the quasi-metallic phase and the effective mobility edge disappears. In this regime, analytical results based on a nonlinear approximation fit well with the numerical simulations. We start by describing the speckle disorder, before discussing the nonlinear approximation and the results obtained.

\section{Speckle Disorder}

Speckle patterns can be identified by their characteristic statistical properties. Indeed, speckles have Fourier components that are limited to a defined spatial frequency range of $2k_{c}$. In addition, the speckle intensity probability distribution obeys a negative exponential statistics \cite{a11}. Experimentally, effective 1D speckle disorder, described extensively in \cite{a15}, is obtained when coherent light is diffracted by a ground glass diffuser with a quasi-1D slit (very large in the y-direction and very small in the x-direction) and focused by a convergent lens. The diffraction pattern is observed in the focal plan. As light is scattered by a rough surface, this induces a random modulation of amplitude and phase of the electric field. The constructive and destructive interferences of the waves originating from different scattering points on the rough surface create randomly distributed light patterns. The field is more strongly diffracted in the x-direction (short direction of the slit) as compared to the other direction. In the y-direction, where the slit's dimension is very large, the field is almost not scattered. In this case the speckle disorder can be considered as almost translationally invariant along the y-direction leading to an effective 1D speckle pattern. In the focal plan $(x,y)$, the electric field amplitude $A(x)$ is a sum of independent complex random variables, corresponding to the scattered field interfering at point $x$. The generated field follows a complex Gaussian distribution. If this diffused light is coupled to an ensemble of cold atoms, it creates an effective random potential determined by $|A(x)|^2$. The random potential can then be written as
 \begin{equation}
V(x)=V_{R}\left( \frac{\left\vert A(x)\right\vert ^{2}}{\left\langle
\left\vert A\right\vert ^{2}\right\rangle }-1\right),
\end{equation}
where $V_{R}$ represents the amplitude of the disorder. In general, the electric field $A(x)$ is a filtered Gaussian process due to a convergent lens, the finite size of the diffuser and the finite  dimension of the 1D-slit \cite{a15,a16}. The electric field can then be described by \cite{a16b}
\begin{equation}
A(x)=\digamma ^{-1}\left( W\digamma (a)\right),
\end{equation}
where $a$ is a Gaussian random variable,  $W$ describes the filter function and $\digamma$ is the Fourier transform. The filter has a spatial cut off frequency $\pm k_{c}$ so that Fourier components of the electric field $A$ vanish for any spatial frequency outside that range. Consequently, the high spatial frequency components of the potential $V$ are zero. This defines the correlation length of the speckle potential, which is given by $\xi=2\pi k_c^{-1}$. The 1D speckle potential obtained this way has a truncated negative exponential distribution \cite{a17}
\begin{equation}
P[V(x)]=\frac{\exp [-\left( \frac{V(x)}{V_{R}}+1\right) ]}{V_{R}}\Theta
\left( \frac{V(x)}{V_{R}}+1\right),
\end{equation}
where $\Theta $ is the Heaviside step function.

\subsection{Numerical generation of the speckle potential}
To generate speckle potentials numerically, we generate a uniform random variable $u_{n}\in \left[ -1,1\right] ,$ and
determine the discrete Fourier transform
\begin{equation}
U_{k}=\overset{L}{\underset{n=1}{\sum }}e^{ikn}u_{n},
\end{equation}
where $L$ is the length of the system. The inverse discrete Fourier transform for the series $U_{k}$ filtered by a square filter $W(k)$ is then
\begin{equation}
u(x) =\frac{1}{\pi}\overset{k_{c}}{\underset{k=-k_{c}}{\sum }}e^{-ikx}U_{k}dk,
\end{equation}
with $dk=\pi/L$ and lattice constant $dx=1$ ($x$ is now an integer). The discrete speckle potential is obtained by evaluating
\begin{equation}
V(x) =V_{R}(\vert u(x)\vert
^{2}-\langle \vert u\vert ^{2}\rangle)
\label{speckle}
\end{equation}
This leads to the following auto-correlation function of the potential:
\be
\langle V(x+y)V(x)\rangle =V_R^2\left(\frac{\sin(2\pi y/\xi)}{2\pi y/\xi}\right)^2
\ee

The speckle potential can take on two different signs of detuning, depending on the sign of $V_R$. For $V_R>0$ the potential is blue detuned and the potential is bound at the bottom, while for $V_R<0$ the potential is red detuned and bound at the top. The speckle potential is strongly asymmetric and we will consider both cases in what follows, since experimentally, both are relevant.

\begin{figure}
	\includegraphics[width=\columnwidth]{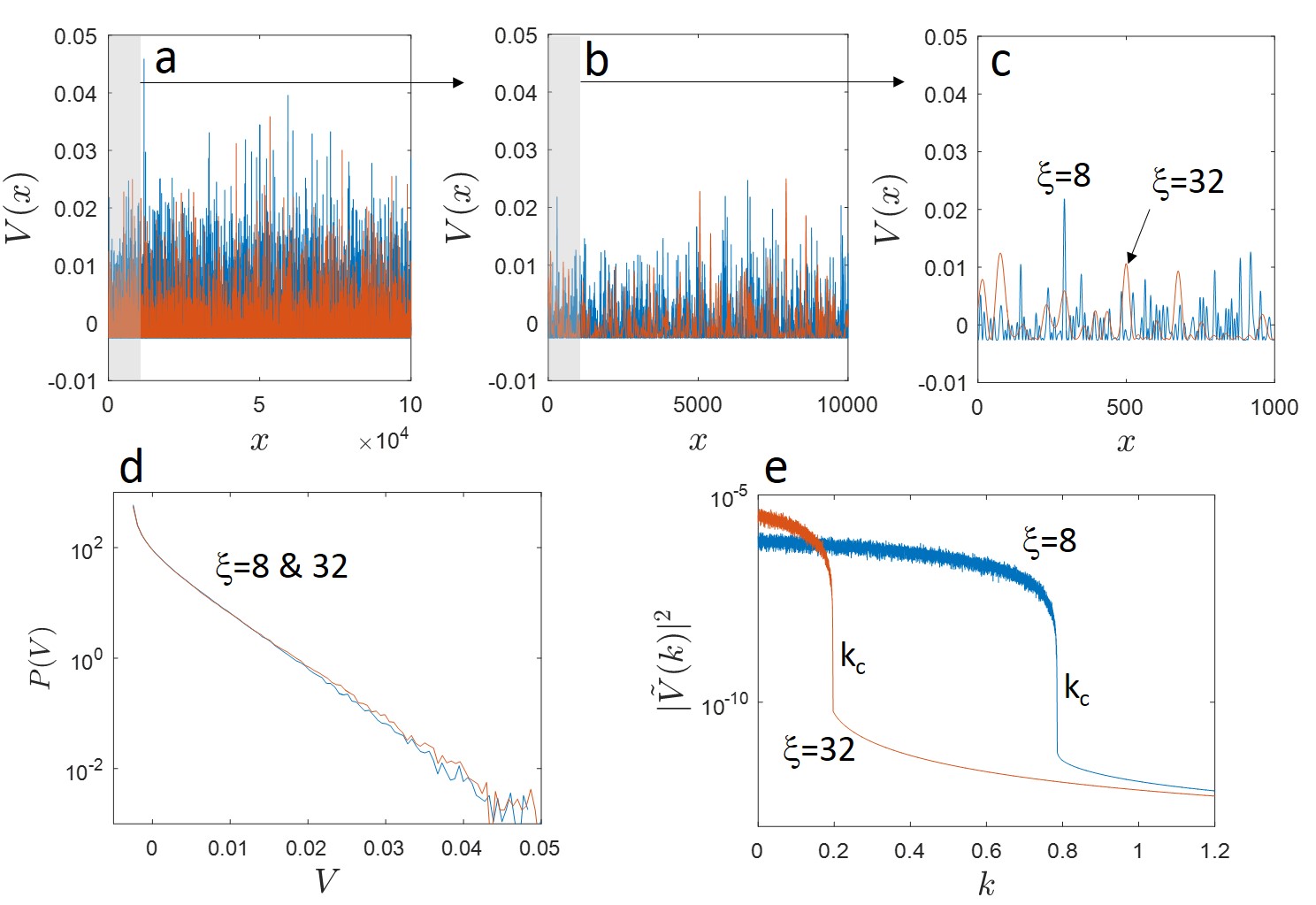}%
	\caption{\label{potential}Example of a blue detuned speckle potential ({\bf a},{\bf b} and {\bf c}) for two different correlation lengths $\xi$, while zooming in from {\bf a} to {\bf c}. In {\bf d}, the distribution of the potential is shown. In {\bf e} the squared Fourier transform is shown for two different correlation lengths, corresponding to different cut-offs. The saturation of $\tilde{V}(k)$ beyond $k_c$ is due to the finite length of the system considered (we used a length of $L=10^5$ here and in the remainder of this work). }
\end{figure}

A numerical speckle potential is shown in figure \ref{potential}, which illustrates the exponential distribution as well as the Fourier transform of the potential. The form of the Fourier transform is responsible for the existence of the effective mobility edge at small disorder, since large components of the spatial frequency vanish, which implies delocalization at large energies (small wavelengths). Equivalently, for fixed energy, delocalization also appears as a function of an increased correlation length since the critical spatial frequency $k_c$ decays with $\xi^{-1}$.

\section{Schr\"odinger solution to the Speckle disorder potential}
The 1D wave equation (or Schr\"{o}dinger equation with $\hbar=2m=1$) is given by
\be
[\partial_x^2+p(x)^2]\psi (x)=0,
\label{schroedinger}
\ee
with classical momentum
\be
p(x)\equiv\partial_x P(x)=\sqrt{E-V(x)},
\ee
where we have defined $P(x)$ as the integrated momentum.
\subsection{Weak disorder approximation}

\begin{figure}
	\includegraphics[width=\columnwidth]{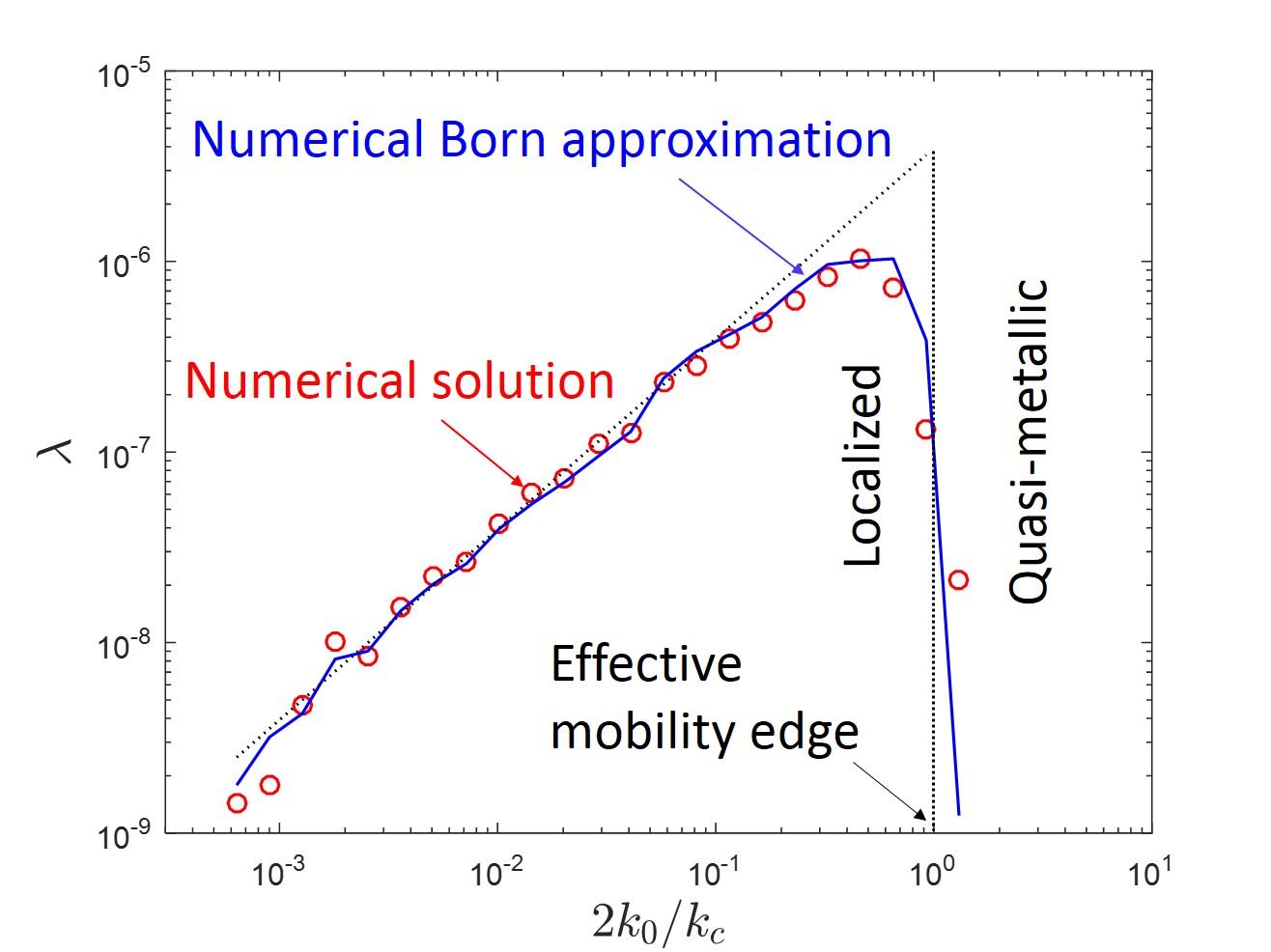}%
	\caption{\label{LowDisorder}Diagram of the low disorder localization behavior as a function of inverse critical spatial frequency. For the Born approximation we use $\lambda_{\tilde{V}}$ obtained from the Fourier transform of the potential (equ. \eqref{lamV}). The analytical expression is given by the dotted line (equ. \eqref{LamAna}), while the numerical result (averaged numerical Fourier transform of the potential) is represented by the full (blue) line. $\lambda_\psi$ is obtained numerically from the wave solution (equ. (\ref{numericalwave})). The numerical parameters are $E=10^{-6}$ and $V_R=-2\cdot10^{-7}$ (red detuned). Here $k_0=\sqrt{E}$ is kept fixed, while $k_c$ is varied and we used an average over 200 different disorder configurations for the numerical results (which leads to small residual fluctuations in the numerical solutions, notably at very large localization lengths).}
\end{figure}

In the case of weak disorder it is possible to find the solution from the Born approximation or equivalently from Fermi's golden rule, since in 1D the localization length is the scattering length. In this case, we have

\be
\lambda_{\tilde{V}}=\frac{|\tilde{V}(2k_0)|^2}{8EL},
\label{lamV}
\ee
where $k_0=\sqrt{E}$ and the Lyapounov exponent $\lambda=L_c^{-1}$ is the inverse of the localization length \cite{izrailev}. For speckle disorder, this result implies that $\lambda_{\tilde{V}}$ vanishes for $2k_0>k_c$, where $k_c$ is referred to as an effective mobility edge \cite{a17,billy}. The result of the Born approximation is shown in figure \ref{LowDisorder}, where the analytical expression is simply given by
\be\lambda_{\tilde{V}}=\frac{\xi V_R^2}{8k_0}\Theta(2k_0-k_c)\label{LamAna}\ee
and shown by the dotted line in figure \ref{LowDisorder}.

The weak disorder case nicely illustrates this effective mobility edge, where there is a sharp drop of the Lyaponov exponent for $2k_0>k_c$. The slight rounding, of the numerically evaluated Born approximation around the transition, is due to the same finite size effect as seen in the Fourier transform shown in figure \ref{potential}.The  existence of a mobility edge can seem surprising at first, since it is commonly accepted that in the presence of even weak disorder all states are localized in 1D. However, this is only the case for uncorrelated disorder. Indeed, correlations in the disorder can lead to extended states in 1D tight binding and continuous potentials \cite{flores1} as well as in 2D \cite{hilke}. A disorder speckle potential is another example of such a correlated potential.

Going beyond the Born approximation, Lugan et al. \cite{lugan} have shown that the effective mobility edge weakens at larger disorder. They used perturbation theory beyond the Born approximation and found a good agreement with numerical data for stronger disorder. The goal here is to go beyond perturbation theory in order to evaluate the localization properties at even higher disorder strengths, in particular, very strong disorder on the "quasi-metallic" side. This regime is highly relevant to experiments on many body localization, which requires strong disorder. Therefore, fully understanding the non-interacting case at large disorder is crucial. Our approach here is based on a new non-linear approximation, which was recently shown to describe Anderson localization for a wide range of disorder strengths \cite{bib1d}.

\subsection{Non linear solution}

To solve the wave equation we look for a solution of the form
\be
\psi(x)=e^{i(P+N)},
\label{pplusn}
\ee
where $N(x)$ is the correction term to be determined. Inserting $\psi$ into (\ref{schroedinger}), we have
\be
i(\partial_xN)^2+2ip\partial_xN+\partial_xp+\partial_x^2 N=0
\ee
In the ERS approximation \cite{Ers}, the idea is to neglect the correction ($N$) to second order
\be
(\partial_xN)^2=0.
\label{ERSapprox}
\ee

An alternate view, is based on solving an analogue non-linear Schr\"odinger equation, where the newly introduced non-linear term disorder averages to zero \cite{bib1d}:
\be
[\partial_x^2+p^2-[\psi^{-1}(-i\partial_x-p)\psi]^2]\psi=0.
\label{schroedingerNL}
\ee
By using equ. (\ref{pplusn}) the last expression leads to
\be
2ip\partial_xN+\partial_xp+\partial_x^2 N=0,
\label{ERS}
\ee
which is equivalent to using approximation (\ref{ERSapprox}). The non-linear approximation corresponds to neglecting the difference between the classical and quantum momentum to second order: $((p+i\partial_x)\psi)^2\simeq 0$. This term vanishes with disorder averaging \cite{bib1d}. The differential operator corresponding to equ. (\ref{ERS}) can be written as
\ba
H_N&=&2ip\partial_x+\partial_x^2\nonumber\\
&=&e^{-2iP}\partial_x(e^{2iP}\partial_x ),
\label{1D}
\ea
where we need to solve
\be
H_N N=-\partial_xp=-p'(x).
\ee
The solution can be obtained by integration, i.e.,
\be
\partial_x N(x)=-e^{-2iP(x)}\int^x e^{2iP(x')}p'(x')dx'.
\ee
Performing the average over disorder yields
\be
\langle f(x) \rangle=\langle p(x) \rangle-\int_{x_0}^xe^{-2ik_0y}c_p(y)dy,
\label{f}
\ee
where we have defined $P+N=\int_{x_0}^xf(x')dx'$ ($x_0$ defines the boundary condition) and where
\be
c_p(y)=\langle k'_v(0) e^{-2i\int_{0}^yk_v(x)dx}\rangle
\ee
is the correlation function of the speckle disorder potential. The average momentum is defined as ($k_0=\langle p(x)\rangle$), its variation from the mean ($k_v(x)=p(x)-k_0$) and its spatial derivative ($k_v'(x)$). To obtain equ. (\ref{f}) we assumed that the disorder is translationally uniform. The decay of the wavefunction (Lyapounov exponent, $\lambda$) is then simply given by $\lambda_{NL}=\Im \langle f(x) \rangle$, which was derived earlier and was shown to be valid for a wide range of disorder strengths \cite{bib1d}.

\section{Numerical method}
To solve the Schr\"odinger's equation numerically, we simply discretize the wavefunction $\psi$ and find the solution by iteration of the following iterative equation:
\be
\psi(x+1)+\psi(x-1)-V(x)\psi(x)=(-E+2)\psi(x)
\ee
and fix the boundary condition at the end of the chain, by taking $\psi(L)=1$ and $\psi(L+1)=e^{ik_0}$. The numerical Lyapounov exponent is obtained by computing
\be
\lambda_\psi=\lim_{L\rightarrow\infty}-\frac{1}{L}\ln|\psi(0)|= -\frac{1}{L}\langle\ln|\psi(0)| \rangle.
\label{numericalwave}
\ee

We also compute the Lyapounov exponent using the non-linear approximation (and boundary condition at $x_0=L$), which can be obtained by solving numerically

\be
\left\{\begin{array}{rl}\theta(x)=&-\int_x^Lk_v(x')dx'\\
	f(x)=&[k_0+k_v(x)]\\
	+&e^{-2i[k_0x+\theta(x)]}\int_x^Lk_v'(x')e^{2i[k_0x'+\theta(x')]}dx'\\
	\psi(x)=&e^{-i\int_x^Lf(x')dx'}
\end{array}\right.
\ee
The corresponding Lyapounov exponent is then obtained by taking the imaginary part of the spatial and disorder average
\be
\lambda_{NL}=\Im \frac{1}{L}\int_0^L\langle f(x) \rangle dx,
\ee
The system size needs to be longer than the correlation length of the disorder potential, which is proportional to $\sim k_c^{-1}$. Hence, we always used $L\gg k_c^{-1}$ when computing $\lambda_{NL}$. Numerically, we used $L=10^5$ for all the figures. Computing the speckle disorder potentials (equ. \eqref{speckle}) for different correlation lengths and configurations is numerically the most intensive part in the overall numerical procedure. In this work we used an average over 200 different disorder configurations.

\subsection{Weak Disorder}

\begin{figure}
	\includegraphics[width=\columnwidth]{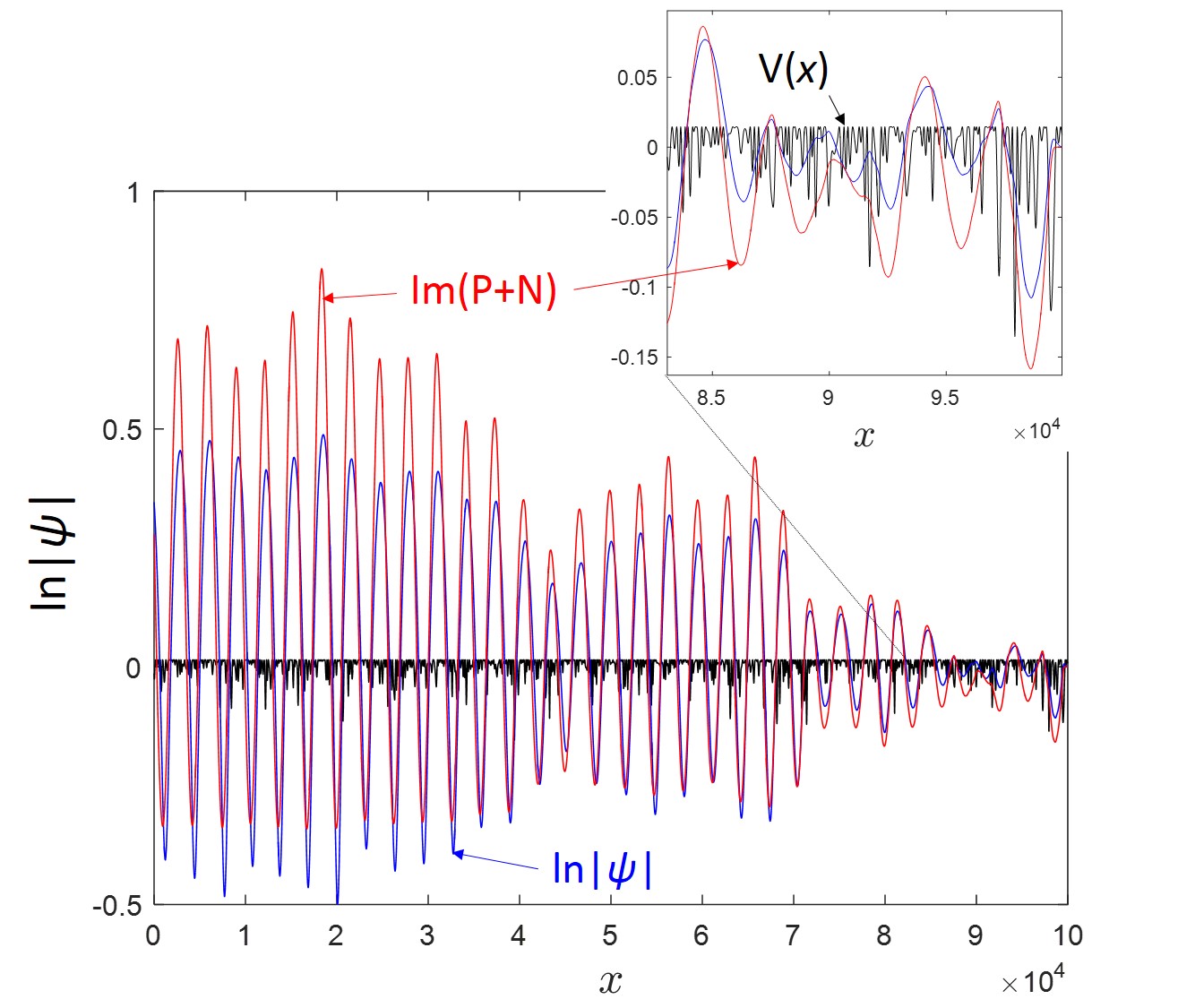}%
	\caption{\label{smo512L2}The potential (multiplied by $10^5$) is shown in black, while the exact numerical solution $\ln|\psi|$ is shown in blue and $\Im(P+N)$ is shown in red. The inset is simply a zoom-in of the right region. Here $\xi=258$ and the wavelength is $2\pi/k_0=6300$ ($E=10^{-6}$).}
\end{figure}

To understand why the Lyapounov vanishes when $\xi\rightarrow0$ (or $2k_0\ll k_c$) as seen in figure \ref{LowDisorder}, it is interesting to analyze the wave solution in this regime shown in figure \ref{smo512L2}. For small correlation lengths as compared to the wavelength, the wave solution averages over the potential fluctuations that oscillate much faster than the wavelength. We can also see how the approximate non-linear solution closely follows the exact numerical solution in this regime. This regime is quite different to the case where $\xi$ is comparable to the wavelength.

\begin{figure}
	\includegraphics[width=\columnwidth]{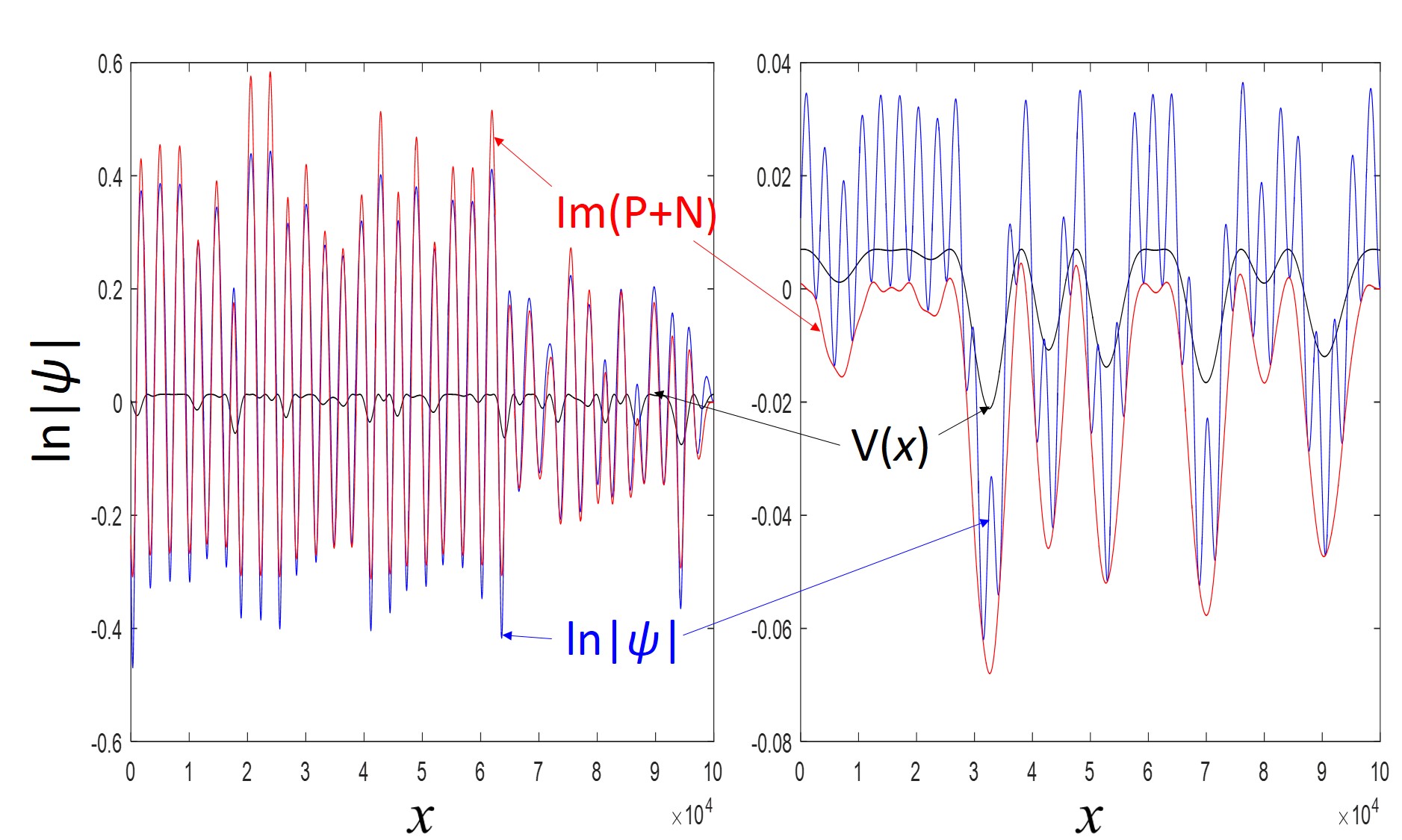}%
	\caption{\label{smo2048_8192L}The potential (multiplied by $10^5$) is shown in black, while the numerical $\ln|\psi|$ is shown in blue and $\Im(P+N)$ is shown in red. In the left graph we have $2k_0/k_c=0.65$, while in the right graph we have $2k_0/k_c=2.6$ and $E=10^{-6}$ in both cases.}
\end{figure}

Here, when $2k_0\lesssim k_c$, as illustrated in figure \ref{smo2048_8192L}, the effect of the speckle potential on the wave solution is strongest, leading to a maximum in the Lyapounov exponent due to the maximized wave interference.

However, beyond the effective mobility edge ($2k_0>k_c$), the situation changes quite dramatically and localization is very weak ($\ln|\psi|$ changes much less). In fact, the envelope of the wavefunction roughly follows the potential as shown in figure \ref{smo2048_8192L} and no interference is occurring. The solution is well described semi-classically, leading to a vanishing Lyapounov exponent (for small disorder). In this regime, the non-linear approximation has similarities with the semi-classical WKB approximation, which neglects the additional resonances due to multiple reflections between local extremas.

In all the cases discussed above, the non-linear approximation closely follows the exact numerical solution. These were all weak disorder cases, where the Born approximation is accurate. Not surprisingly, the non-linear approximation  converges to the Born approximation at small disorder. The more interesting situation arises when disorder is increased. Here the Born approximation breaks down for speckle potentials \cite{lugan}. An obvious difficulty arises when the potential exceeds the energy locally. This will happen quite naturally for a blue detuned speckle potential, where $V_R>0$, since in this case, there is no upper bound on the value of the potential. If the probability of such a situation ($V>E$) is small (exponentially small for weak disorder), the results will not be much affected. This is in fact the case for the regime where $\xi$ is small, since in this case the support where $V>E$ is of vanishing measure. Hence, even for $V_R\simeq E$, the Born approximation is still good as long as $k_0\ll k_c$. We indeed observe the characteristic linear behavior as a function of the correlation length ($\lambda\sim V_R^2\xi$) in figure \ref{LowDisorder}. These results apply for both signs of the speckle potential ($V_R>0$ and $V_R<0$).

\subsection{Strong Disorder}
\begin{figure}
	\includegraphics[width=\columnwidth]{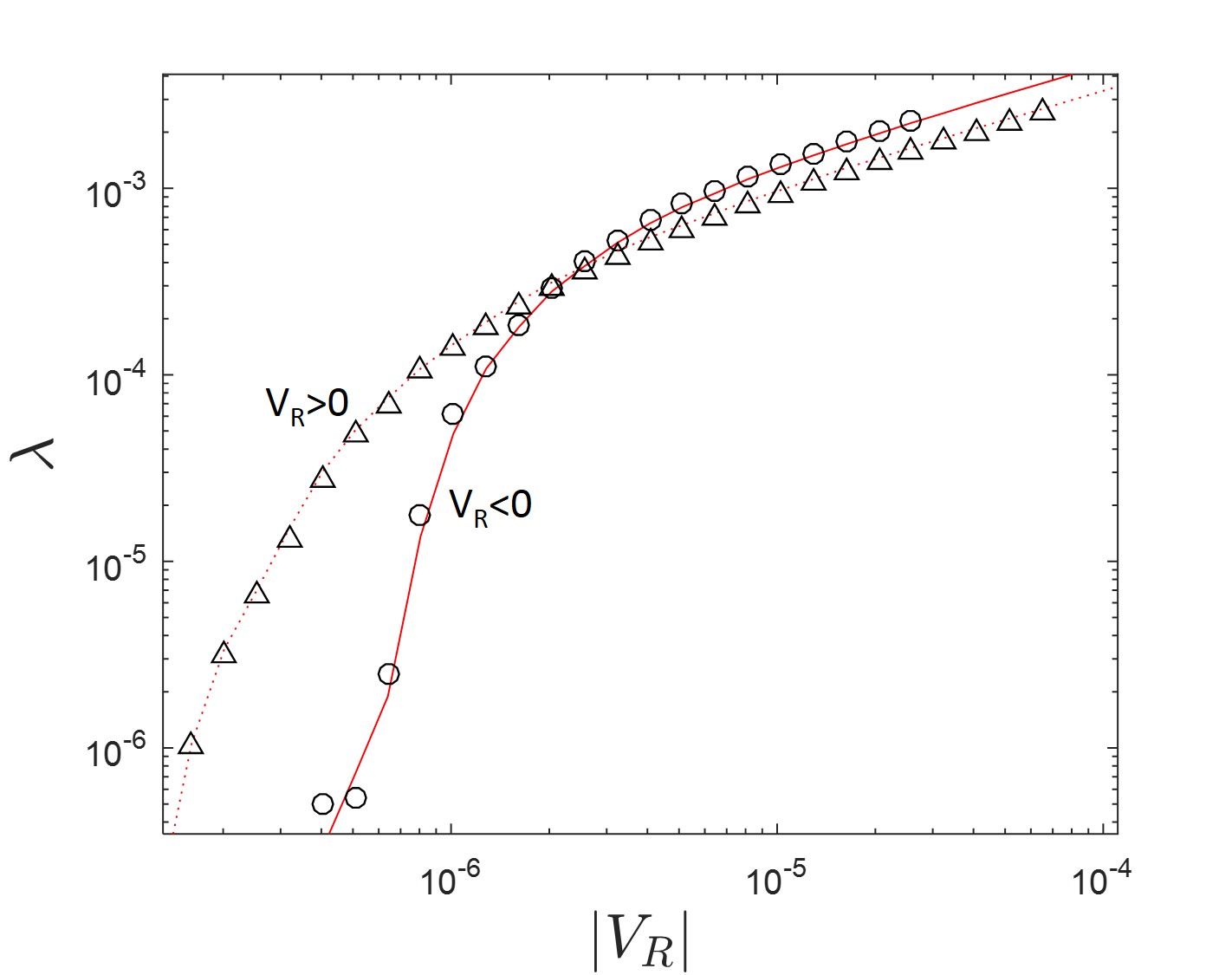}%
\caption{\label{disorderdep} Lyapounov exponent as a function the disorder amplitude for both blue and red detuned speckle potentials. The open symbols correspond to $\lambda_{NL}$, while the lines correspond to $\lambda_\psi$. The numerical parameters are $2k_0/k_c=2.6$ with $E=10^{-6}$. }
\end{figure}

The situation changes dramatically for the "quasi-metallic" regime ($2k_0>k_c$). Here the effect of strong disorder is non-trivial. This is the regime that we will now focus on. The Born approximation would simply yield $\lambda_{\tilde{V}}=0$ regardless of the disorder strength. We show in figure \ref{disorderdep} how the Lyaponov exponent increases with disorder strength when $2k_0>k_c$.

There is a strong asymmetry between the blue and red detuned potentials. This asymmetry is due to the different probabilities of having points where $E<V$. This asymmetry also exists for the onset of localization as discussed in ref. \cite{delande}. The disorder dependence is closely described by our non-linear formalism as shown in figure \ref{disorderdep}, which shows $\lambda_\psi\simeq\lambda_{NL}$ for all disorder strengths. This validity extends all the way to very strong disorder ($V_R \gg E$), where $E<V(x_{sub})$ over a large subset $x_{sub}$.

\begin{figure}
	\includegraphics[width=\columnwidth]{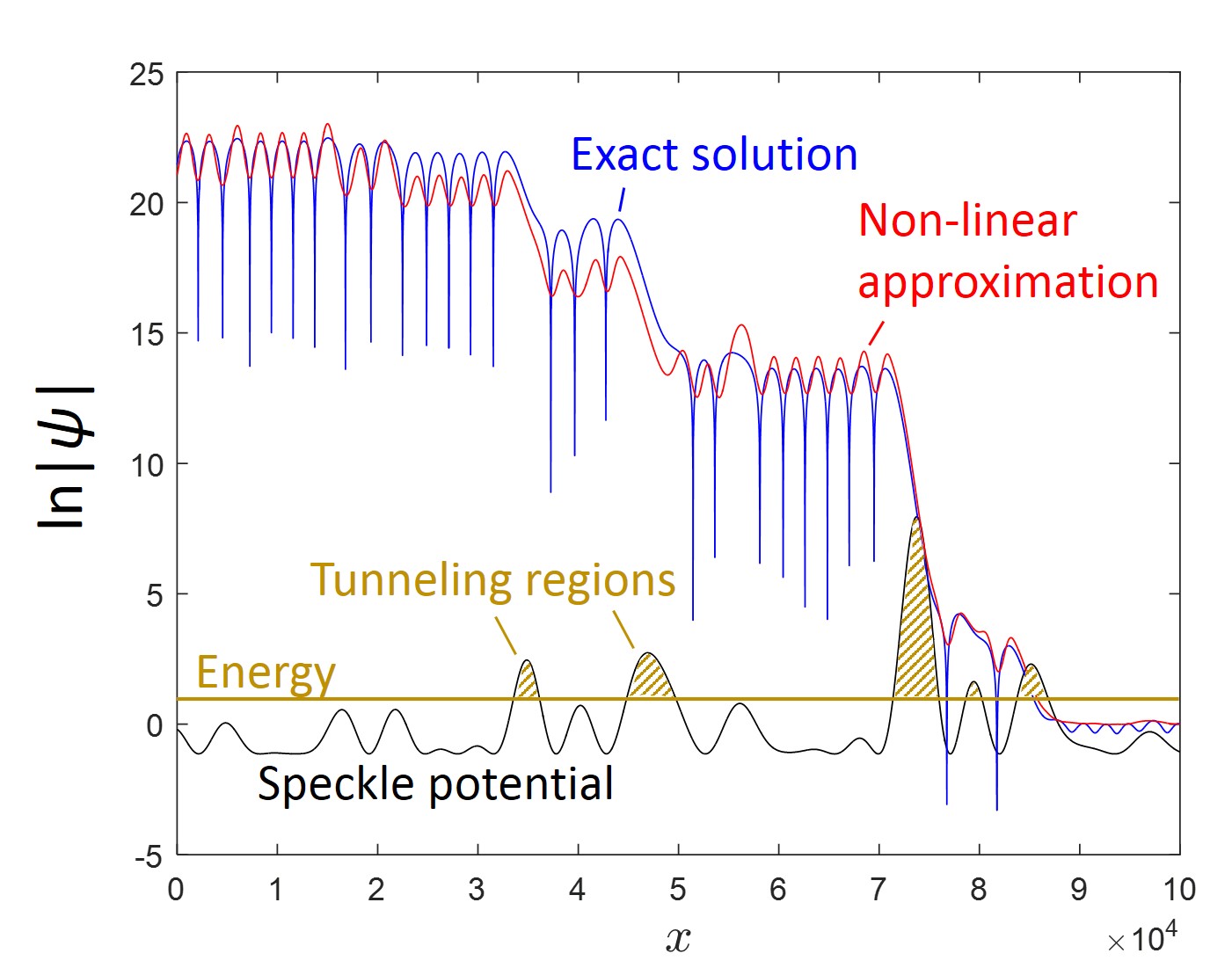}%
	\caption{\label{StrongDis} The potential (multiplied by $10^5$) is shown in black, while the exact numerical solution $\ln|\psi|$ is shown in blue and $\Im(P+N)$ is shown in red. The tunneling regions ($E<V$) are indicated in brown. The numerical parameters are $E=10^{-6}$, $V_R=2.5\cdot 10^{-6}$, and $\xi=8\cdot 10^{3}$.}
\end{figure}

It is instructive to inspect the wave solution in the very strong disorder regime, which is shown in figure \ref{StrongDis}. This figure nicely illustrates the effect of the potential maxima, where the wave solution has amplitude jumps due to tunneling regions. The number and size of these events largely define the Lyapounov exponent.

\begin{figure}
	\includegraphics[width=\columnwidth]{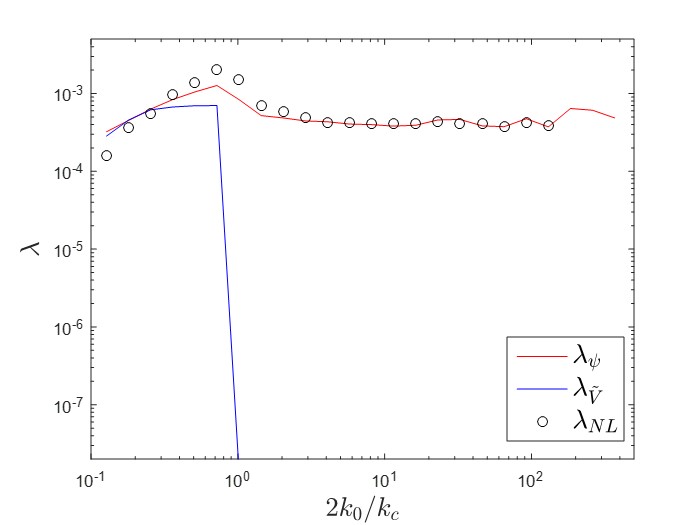}%
	\caption{\label{kcdep} Lyapounov exponent as a function of $k_c^{-1}$ with $E=0.04$ and $V_R=0.006$). These parameters illustrate how strong disorder destroys the "weak disorder effective mobility edge" over a wide range of correlation lengths. Shown are the Lyapounov exponents obtained numerically using equ. \eqref{numericalwave} for $\lambda_\psi$,  the non-linear approximation using \eqref{f} for $\lambda_{NL}$, which provide an excellent agreement as opposed to the Born approximation ($\lambda_{\tilde{V}}$).}
\end{figure}

We now look at the behavior of the Lyapounov exponent for strong disorder as a function of $2k_0/k_c$. In contrast to the weak disorder result, which exhibits an effective mobility edge, in the presence of strong disorder, there is only a weak dependence on $k_c^{-1}$ as show in figure \ref{kcdep}. While there is still a small local maximum close to $2k_0\simeq k_c$, $\lambda$ is almost to constant over a very wide range of $k_c$ for $2k_0>k_c$. This has to be contrasted to the behavior when $2k_0<k_c$, where $\lambda\sim k_c^{-1}$ (while keeping $k_0$ constant). Strong disorder suppresses the effective mobility edge, due to the increase in  tunneling regions. Remarkably, our non-linear approach and its expression for the Lyapounov exponent, allows us to accurately determine the localization behavior even in the very strong disorder case when $E<V$ in many regions.

\section{Summary}

Summarizing, we have evaluated the correlation length ($\xi$) dependence of localization in speckle disorder for all disorder strengths. At weak disorder the Lyapounov exponent is maximum close to $2k_0\simeq k_c$ (or the wavelength $\simeq 2\xi$) assuming a constant energy and disorder amplitude. This behavior is well described by the Born approximation, which accurately describes the effective mobility edge. However, when the disorder is increased, the localization length remains roughly constant when $k_0>k_c$ implying the disappearance of the effective mobility edge. These results show that at large speckle disorder all states are strongly localized, regardless of the spatial frequency associated with the speckle potential. This behavior can be calculated in terms of a non-linear approximation, which is non-perturbative in the disorder strength, allowing us to express the Lyapounov exponent in terms of a correlation function of the disorder potential for any disorder, even when $E<V$ in some regions. The strong disorder results have important implications for many-body localization (MBL), since single particle localization, which is a requirement for MBL, exists regardless of the correlation length, as long as the disorder is strong enough.

We would like to thank Alain Aspect and Laurent Sanchez-Palencia, who inspired us to work on this problem.

\end{document}